# Attraction of Meso-Scale Objects on the Surface of a Thin Elastic Film Supported on a Liquid

Aditi Chakrabarti* and Manoj K. Chaudhury
Department of Chemical Engineering
Lehigh University, Bethlehem, PA 18015

**ABSTRACT:**

We study the interaction of two parallel rigid cylinders on the surface of a thin elastic film supported on a pool of liquid. The excess energy of the surface due to the curvature of the stretched film induces attraction of the cylinders that can be quantified by the variation of their gravitational potential energies as they descend into the liquid while still floating on the film. Although the experimental results follow the trend predicted from the balance of the gravitational and elastic energies of the system, they are somewhat underestimated. The origin of this discrepancy is the hysteresis of adhesion between the cylinder and the elastic film that does not allow the conversion of the total available energy into gravitational potential energy as some part of it is recovered in stretching the film behind the cylinders while they approach each other. A modification of the model accounting for the effects of adhesion hysteresis improves the agreement between theoretical and experimental results. The contribution of the adhesion hysteresis can be reduced considerably by introducing a thin hydrogel layer atop the elastic film that enhances the range of attraction of the cylinders (as well as rigid spheres) in a dramatic way. Morphological instabilities in the gel project corrugated paths to the motion of small spheres, thus leading to a large numbers of particles to aggregate along their defects. These observations suggest that a thin hydrogel layer supported on a deformable elastic film affords an effective model system to study elasticity and defects mediated interaction of particles on its surface.

*E-mail: adc312@lehigh.edu.

## 1. INTRODUCTION

Interaction of particles on the surface of a liquid mediated by the joint effects of capillarity and gravity is a well-studied problem[1-6]. Elasticity mediated interactions of molecules and particles on a thin film or a solid surface has also been discussed extensively in the literature[7-11]. The ability to manipulate the properties of soft materials has opened up new experimental and theoretical studies in this subject in recent years involving liquid crystals[12-15] and gels[16-17]. We recently reported[16,17] interactions of different types of solid beads in an ultra-soft gel as a function of latter's elasticity (shear modulus $\mu$). If the density of the bead ($\rho_b$) is much larger than that of the gel ($\rho_g$) and if its size (radius $R$) is significant, it plunges inside the gel and becomes neutrally buoyant due to elastic deformation forces in the gel. While still submerged, the beads, however, attract each other due to the combined effects of elastic and the surface forces of the gel[16]. On the other hand, if its density is not significantly larger than the gel, the beads float on its surface again mainly by the force due to elastic deformation of the gel but attract[17] each other somewhat like the hydrophobic particles do on the surface of an ordinary liquid[1-6]. Whether the bead will sink in the gel or float on it can be discerned on the basis of a dimensionless number [$EBo = (\rho_b - \rho_g)Rg/\mu$, $g$ being the acceleration due to gravity] – an elastic *Bond* number -- in analogy to the classical *Bond* number, a large value of which indicates sinking, while a low value implies floatation. While studying the interactions of particles in a gel is both interesting and useful, analyses of these interactions are somewhat complicated because of various non-linear effects intrinsic to the gel and the phenomena involved, the full understanding of which requires a 3d analysis of the non-linear field equations describing the deformation of the gel. A model system that simplifies the role of elasticity in particle interactions is a two dimensional elastic membrane that undergoes large deformation under stretching. The situation can be simplified further by studying interactions of two parallel cylinders, which, is ideally a 2d problem. These studies are also valuable in the context of understanding the elasticity mediated interaction of particles that is generic to various phenomena involving biomolecules on a cell membrane.[8-9] Here we report interactions of particles on a thin elastomeric film that is supported on the surface of an incompressible liquid. The advantage of this system is that the role of the thin film can be studied explicitly while the hydrostatic pressure in the liquid tempers the length scale over which interaction prevails. Superficially, the elastic Bond number [$EBo = (\rho_b - \rho_l)R^2 g / T$, $\rho_l$ being the density of the supporting liquid], in this case, resembles the classical Bond number, except that the film tension ($T$) here is composed of the surface free energies ($\gamma$) of the air-film and film-liquid interfaces as well as an elastic tension $T_E$. Furthermore, as there is no pre-existing tension in the film, the elastic tension here is encumbered by the strain in the film induced by its stretching imposed by the weights of the cylinders. Nevertheless, by varying the value of $T_E$, the magnitude of the vertical penetration of the cylinder in the liquid while still floating the membrane and the range of interaction can be easily controlled.

The main experiment involves the interactions of two parallel cylinders on the surface of a thin elastomeric (polydimethyl siloxane) film supported on the surface of a mixture of glycerol and water. As mentioned above, we chose cylinders as opposed to spheres mainly because of the simplicity of data analysis, in that energy minimization in 2d suffices to capture the main physics of such interactions. Because of the excess elasto-capillary field energy on the surface of the film, the cylinders roll towards each other and come into close contact. At this juncture, we point out that these types of experiments are not easy to perform on a liquid surface as such a parallel configuration is intrinsically unstable; thereby the cylinders approach each other displaying various metastable configurations with a non-parallel geometry[2]. In the case with solid cylinders on an elastic film, the sliding friction at the interface stabilizes their parallel configuration from a modest distance all the way to contact. These interactions are the combined effects of elasticity and gravity. As the elastic strain energy of the deformed film is released, the cylinders descend further in the liquid while still supported by the elastic film. Estimation of the change in the gravitational potential energy, therefore, provides a first order estimate of the energy of interaction of the cylinders as we have also shown recently with various particles interacting in a gel. During the course of the analysis of the data, we noted that the experimentally observed energy of interaction is somewhat smaller than that predicted theoretically. We argue that the discrepancy is related to the hysteresis of adhesion[18,19] between the cylinder and the elastic film. As a cylinder rolls, its leading edge makes new contact with the film, whereas the contact is broken at its rear edge. Due to the difference of these two adhesion energies, some of the elastic field energy released from the intervening region of the cylinders is used up in further stretching the film behind the cylinders. A simple experiment in which a cylinder was forced to roll on the surface of the elastic film provided ample evidence that the role of adhesion hysteresis cannot be ignored. If this frictional resistance is reduced by supporting a layer of an ultra-soft hydrogel on the surface of the elastic film, cylinders as well as spherical beads attract each other from a much larger distance than what is observed with an elastic film alone.

## 2. RESULTS AND DISCUSSION

**2.1. Estimation of Energy of Attraction Using Gravity.** Both the stability of the cylinders in the vertical direction as well as their mutual attraction are the results of the balance of the elastic stretching energy of the film, and the gravitational potential energies of the deformed liquid as well as that of the cylinder itself. For the case of a single cylinder, the profile of the deformed surface can be obtained from the usual minimization of a functional comprising of the gravitational potential energy of the liquid and the stretching energy of the film (see the appendix for a detailed discussion of equation 1):

$$U_1 = \frac{L}{2}\int \rho g \xi^2 dx + \frac{L}{2}T\int \xi_x^2 dx \tag{1}$$

Where, $\xi(=\xi(x))$ is the deformation of the surface measured from the far field undisturbed surface of the film, $\rho$ is the density of the liquid, $g$ is the acceleration due to gravity and $T$ is the tension in the film. A clarification of equation 1 is waranted here. Equation 1 is what one would expect for a deformed liquid surface as the surface of the liquid has a surface free energy that increases with the curvature of the surface. However, for the elastic film considered here, it has no excess free energy in the unstretched state in the absence of a pre-stress. We confirmed that the pre-stress in the film is negligible from a simple obervation in which a small hole is created at the center of the film with a sharp needle. If there is a substantial pre-tension, one would expect the hole to grow. We, however, observed that the punctured hole either does not grow or grows by such a small amount that it is inconceivable that any substantial pre-tension exists in the films. The justification of using a linear model for elastic modulus is based on the experimental observation that the profile of the membrane on both sides of the cylinders decay exponentially and that in between them it decays following a cosine hyperbolic function (see below).

A minimization of the energy functional [i.e. $\delta U_1/\delta\xi=0$] leads to the familiar[3] differential equation of capillarity, [i.e $\xi_{xx} = \alpha^2\xi$], the solution of which leads to an exponential variation of surface deformation $\xi = \xi_0 e^{-\alpha x}$ with a decay length $\alpha^{-1} = \sqrt{T/\rho g}$. The exponential solution ensures that $\xi$ vanishes far field with its maximum value $\xi_o$ at $x = 0$. The experimental measurement of the deformed surface profile in conjunction with the preceding expression for $\xi$ can be used to estimate the decay length $\alpha^{-1}$. The differential equation needed to estimate the surface profile[20,21] of the film that undergoes both bending and stretching would be: $D\xi_{xxxx} - T\xi_{xx} + \rho g \xi = 0$ ($D$ being the bending constant), which has a periodic solution with an exponential decay when bending dominates. A scaling analysis leads to the ratio of the bending and the stretching terms in the above equation as $(1/\varepsilon)(H\alpha)^2$, where $H$ is the thickness of the film (<10 $\mu m$), $\alpha^{-1}$ (~1 cm) is the relevant lateral length scale (the decay length: see below) and $\varepsilon$ (~$10^{-1}$) is the strain in the membrane due to the stretching induced by the weight of the cylinder. Using the above parameters, the ratio of the bending to stretching terms is on the order of $10^{-5}$; thus, the bending term is negligible as compared to the stretching of the elastomeric membrane. Its neglect is also justified on the basis of the observation that the interface profile is only exponential in $x$ within an observation window of about 5 cm; no oscillatory profile was visible in the region where membrane undergoes out of plane stretching.

In order to estimate the energy of interaction of the parallel cylinders, we need to consider their potential energies in addition to the stretching energy of the film as well as the gravitational free energy of the liquid in the region intervening the cylinders and beyond. The total excess energy $U_T$ then becomes:

$$U_T = -2m^* g h + L\left[\int_0^\infty \rho g \xi^2 dx + T\int_0^\infty \xi_x^2 dx\right] + \frac{L}{2}\left[\int_{-\ell/2}^{+\ell/2} \rho g \xi'^2 dx' + T\int_{-\ell/2}^{+\ell/2} \xi_x'^2 dx'\right] \quad (2)$$

Where, $\ell$ represents the separation distance between the lines where the film meet the cylinders and $h$ represents the depth of immersion of the base of the cylinder measured from the far away undeformed surface of the film (Figure 5 a). The minimizations of the above energy functional with respect to $\xi$ and $\xi'$ lead to two differential equations, the subsequent solutions of which lead to exponential profiles of the film beyond the intervening space, but has a solution of the type shown in equation (3) in the space intervening the cylinders:

$$\xi' = \xi'_0 \frac{\cosh(\alpha x)}{\cosh(\alpha \ell / 2)} \tag{3}$$

Using these surface profiles and taking the depth $h$ to be nearly equal to the $\xi_o$ (based on experimental observations), equation (2) can be integrated to obtain the following result:

$$U_T = -2m^* g \xi_0 + \frac{\rho g \xi_0^2 L}{\alpha}\left(1 + \frac{\sinh(\alpha \ell)}{2\cosh^2(\alpha \ell/2)}\right) \tag{4}$$

The stability condition $\partial U / \partial \xi_0 = 0$, furthermore, leads to:

$$m^* g = \frac{\xi_o \rho g L}{\alpha}\left(1 + \frac{\sinh(\alpha \ell)}{2\cosh^2(\alpha \ell/2)}\right) \tag{5}$$

Substitution of the above expression in $U_T$, and appropriate algebraic manipulations yield the following result:

$$U_T = -m^* g \xi_0 \tag{6}$$

Equation 6 is a form of $U_T$ that can be estimated from the change in the gravitational potential energy of a single cylinder in which $\xi_o$ depends implicitly on the distance of separation $\ell$, which can also be expressed explicitly as a function of $\ell$ using equation 5. Combining these two forms of the energy, we obtain the following equation:

$$\frac{h_\infty - h(\ell)}{h_\infty} = \frac{\sinh(\alpha \ell)}{2\cosh^2(\alpha \ell/2) + \sinh(\alpha \ell)} \tag{7}$$

Since there is not much of a difference in the experimentally measured values of $\xi_0$ and $h$, we express the net change in energy with respect to the final energy in terms of $h$ as it is more convenient to measure the depth of submersion of the cylinder from the images. $h_\infty$ is the final depth of submersion when the cylinders make contact. Thus, from the vertical descents of the cylinders alone, it is possible to estimate the form of the attractive energy of the cylinders as a function of their distance of separation.

**2.2. Thin Elastic Films Supported on a Pool of Liquid.** Thin elastomeric films were prepared by depositing drops of a (1:1) mixture of Sylgard 184 and Sylgard 186 (Dow Corning®) on the surface of a (1:1) Glycerol-Water solution that pre-filled part of a petri dish. Such a composite elastomer possesses high tear strength with an elastic modulus ~ 1MPa. Thus, it is quite durable and withstands the weights (1.2 gm) of the steel cylinders placed above it. The choice of liquid was made on the basis of certain complementary properties it affords. For example, its dispersion component (29 mN/m) of the surface tension being higher than that of PDMS (22 mN/m) allows uniform spreading of the drops of PDMS on its surface without undergoing dewetting that could happen on the surface of pure water. Furthermore, its moderate viscosity (6 cP, 20°C)[22] ensures that the film remains reasonably undisturbed while handling and transporting the samples from one location to another. The uniform interference colors observed on the surfaces of the cured elastic films are indicative of the fact that uniform films of PDMS can be successfully prepared using the mixture of sylgard 184 and sylgard 186 on the water-glycerol solution.

**2.3. Estimation of the Decay Length ($\alpha^{-1}$).**

Our main experiment was to study the distance dependent attraction of two solid cylinders on the surfaces of the thin elastic films with their long axes parallel to each other and to analyze the data in view of the equation (eq 7) proposed as above. These experiments were performed with steel cylinders on the surface of PDMS films of different thicknesses, hence with different tensions. In order to study the universal behavior of these attraction, the non-dimensional descents of the cylinder (LHS of equation 7) were plotted in terms of a non-dimensional distance of separation (

$\alpha \ell$ ). Our first task, therefore, was to estimate $\alpha^{-1}$ as a function of film thickness, which we accomplished in two different ways and selected the more reliable method of estimating the same based on the criteria discussed below.

**A. Estimation of the Decay Length from the Side-View.**

A cylinder was placed on the elastic film parallel to the edge of a square petri dish containing the solution of water and glycerol that supported the film. Image of the side view of the deformed profile was captured by placing the axis of microscope (equipped with a CCD camera) parallel to that of the cylinder, which were then analysed in ImageJ and fitted with an exponential equation of the form:

$$(\xi_0 - \xi) = \xi_0(1 - e^{-\alpha x}) \qquad (8)$$

All the analyses were performed with an Originlab software, by setting the point where the film meets the cylinder as a reference. The decay length $\alpha^{-1}$ for each film could thus be obtained from the fitted profile directly. The decay length $\alpha^{-1}$ could also be estimated from the deformed profile of the elastomeric film in between the parallel cylinders floating on an elastic film in view of equation 3. While both these methods yielded similar values of $\alpha^{-1}$, there are concerns that these values could be somewhat obscured by the folding and wrinkling instabilities (Figure 1) that ensue near the edges of the cylinders (somewhat similar to what happens with a drop liquid on a floating elastic film[23] ), which is the region that is captured by the optical method used here. We thus opted for another direct method of estimating $\alpha^{-1}$ from the deformed profile of the film near the mid-section of the cylinder as discussed below.

**B. Estimation of the Decay Length from the Mid-Section of the Surface Profile.**

In order to visualize the profile of the surface near the mid-section of the cylinder, we deposited a thin line of ink on the film spanning across a square petri dish prior to placing a cylinder upon it. (Figure 2 a) After gently placing a cylinder on the surface of the film perpendicular to the ink line, its deformation was captured with a camera. The deformed profile from the image was then analysed using ImageJ and fitted with equation 8 to obtain the decay length $\alpha^{-1}$ of the film. (Figure 2 b).

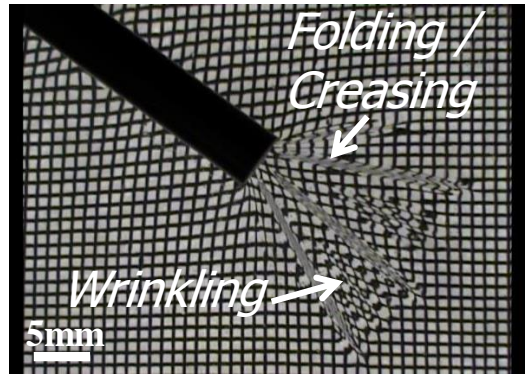

**Figure 1.** Plan view of the folding and wrinkling instabilities at the end of a cylinder placed on an elastic film (3.3 μm thick) supported on a pool of a water-glycerol solution. The wire mesh lined with the base of the petri dish containing the sample shows the deformations in the film surface.

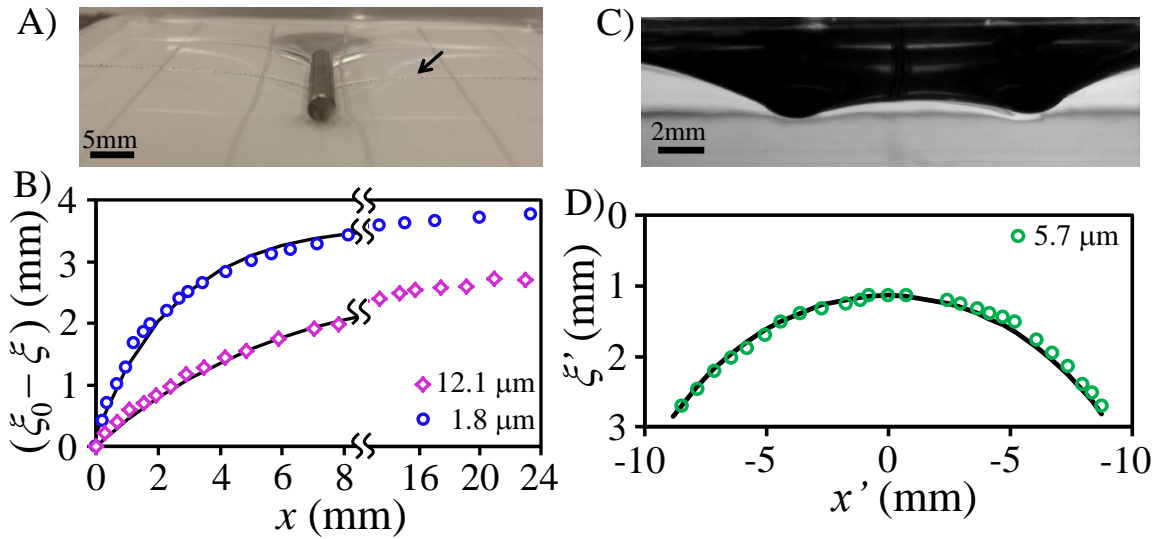

**Figure 2.** (A) The ink line (as shown by the arrow) follows the deformation of the surface of a 7.2 micron thick elastic film supported on the glycerol-water solution when a steel cylinder (diameter 1/8") is placed upon it. (B) The deformed profiles analysed from images of two different elastic films (1.8 μm and 12.1 μm) and the values of $\alpha^{-1}$ obtained from the analysis of the data using equation 8 are 2.7 mm and 5.9 mm respectively. (C) A typical snapshot of the intermediate profile between two steel cylinders on a 6.2 micron thick film. (D) Intermediate profile between two steel cylinders resting on a 5.7 micron film ( having an initial separation distance of 17 mm) fitted with equation 3 to obtain the decay length (5.6 mm).

**Analysis of Data by Contrasting the Two Approaches:**

The values of $\alpha^{-1}$ as a function of the thicknesses of the PDMS films obtained from the above two methods fortunately do not differ in a significant way although a slight difference was observed as noted below. If the surface tension of the solid contributes to the total tension of the film, $T$ should be expressed as $T = T_E + \gamma$, where elastic tension $T_E \sim f(E, H, \varepsilon)$, $E$ being its elastic Young's modulus, $\varepsilon$ is the strain in the film and $H$ is its thickness (see appendix). As the decay length is given by $\alpha^{-1} = \sqrt{(T_E + \gamma)/\rho g}$, we plot the experimental values of $\alpha^{-2}$ as a function of $H$ and found an empirical systematic linear relationship (figure 3). Thus the extrapolated value for $H=0$ should provide an approximate estimate of the surface tension of the solid film, which is contributed by the free surface of the film and that of the film-solution interface. While the decay lengths were measured for a range of elastic film thicknesses, the data were well-behaved (i.e. $\alpha^{-2}$ is fairly linear with $H$) for films of thickness less than 15 μm. For films thicker than 15 μm, some wrinkling was observed underneath the cylinder along its length of contact that was not evident in the thinner films. Since these wrinklings use up some of the available energy, the decay length is somewhat underestimated that introduces uncertainity in quantitative analysis of the attractions of the cylinders on such thicker films. We thus avoided using such thick films for additional measurements and analysis in the current study. At a 95% Confidence Limit, the surface tension $\gamma$ is found to be about $-7 \pm 29$ mN/m from the data obtained from the side view of the profiles and about $56 \pm 27$ mN/m with the profile obtained from the mid section of the cylinder. Although both the methods yield large uncertainity of estimating $\gamma$, the latter method, at least, yields a positive mean value of the surface tension of the solid. Based on the above observations and due to the possibility of $\alpha^{-1}$ measured from the side view being somewhat obscured by the folding and wrinking of the film around the edges, we relied on its value obtained from the profile near the mid-section of the cylinder.

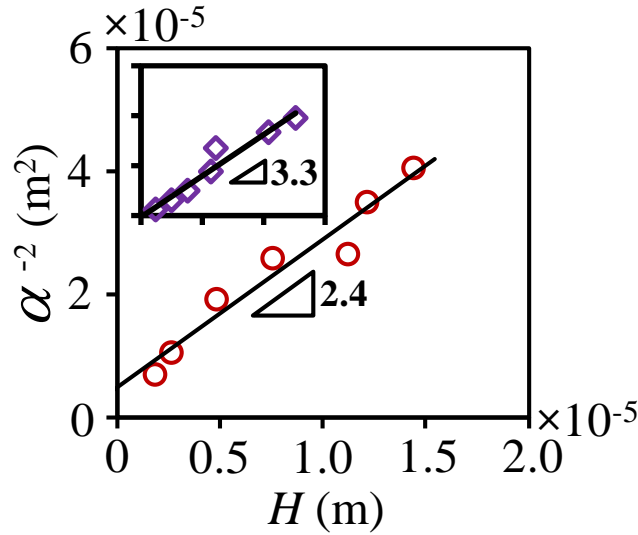

**Figure 3.** Squared values of the decay length $\alpha^{-1}$ are plotted as a function of the film thickness $H$. The red open circles represent the data obtained from the profile of the deformed line of ink as discussed in section 2.3 B. (Inset) The purple open diamonds represent the decay lengths measured from the side view of the deformed profile as discussed in section 2.3 A. The scales of the X and Y axes in the inset graph are same as those of the main plot.

**2.4. Estimation of Elastic Modulus of the Film.**

In order to ascertain that the thin films have been adequately crosslinked on the surface of the water-glycerol solution, we estimated its elastic modulus. For a quick estimation of the elastic modulus, we induced buckling by compressing a section of the film by bringing in closely the edges of two glass slides gently touching a section of the film resting on the pool of a liquid. The Young's modulus $E$ of the film was estimated from its bending modulus $[D = Eh^3/12(1-\nu^2)]$, which is related to the buckling wavelength ($\lambda$) as[20,21] $D = \rho g (\lambda/2\pi)^4$. While the Young's modulus of the PDMS films was in the range of 1-2 MPa, the method was suitable only for thinner films (4-10 $\mu$m). It was rather difficult to generate uniform buckles perperndicular to the edges of the glass that made it difficult to measure the wavelength accurately and hence the method was unsuitable for the thicker films. Since our main purpose was to determine if all the films had the same elastic modulus, we opted for a different technique to test the same by deforming the films by a thin plate as described below.

In this method, a thin plate (Cover Glass Slide, 0.18mm, 24 mm x 50mm ) was pushed into the film vertically and the force of penetration was measured as a function of the displacement of the edge of the plate (figure 4 a). Modification of equation (1) by considering the energy due to the profiles on either side of the plate yields the total gravitational and elastic energies of the liquid and the elastic film as $U_2 = 2U_1$. The force per unit width of the glass cover slip ($F/L$) can be obtained as $F = -\partial U_2 / \partial \xi_0$. Using the method described in the appendix, we obtain two expressions (under two different assumptions) for the pushing force $F$ as a function of the depth of penetration $\xi_0$ as:

$$F/L = 1.3(\rho g)^{3/4}(EH)^{1/4}\xi_0^{3/2} \tag{9A}$$

$$F/L = 1.1(\rho g)^{2/3}(EH)^{1/3}\xi_0^{4/3} \tag{9B}$$

The results summarized in figure 4 b show that $F/L$ is slightly super-linear with respect to $\xi_o$ with a power law exponent close to 1.3, which is closer to the prediction of equation 9B. Using a value of $E$=1.2 MPa, which is the modulus expected[24] of a composite film of sylgard 184 and 186, a plot of $F/L$ against $(\rho g)^{3/4}(EH)^{1/4}\xi_0^{3/2}$ for seven different elastic films (1.8 $\mu$m to 14.4 $\mu$m) yields a slope of $1.58 \pm 0.02$, which is close to the theoretically predicted value 1.3 (equation 9a).

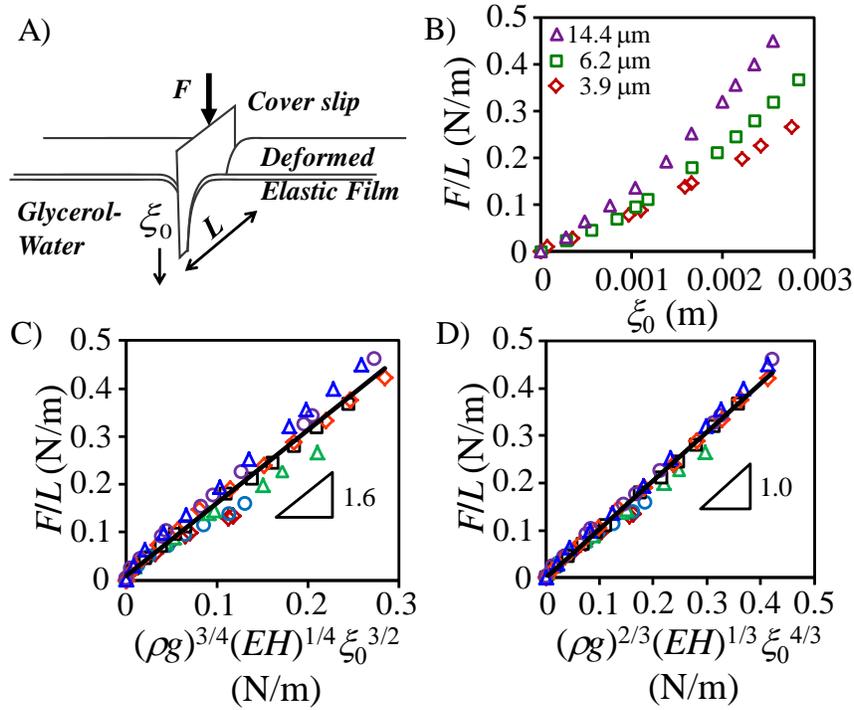

**Figure 4.** (A) Schematic of a thin cover glass (width $L=$ 50mm) indenting the surface of the composite elastic film of Sylgard 184 and 186 due to a vertically applied force $F$. The indentation depth $\xi_0$ increases slightly super-linearly with $F$ (not shown here) (B) The force per unit width of the glass plate $F/L$ increases super-linearly with $\xi_0$ with an exponent close to 1.3. Here data are shown for three representative film (1.8, 6.2, 14.4 $\mu$m). (C) $F/L$ is plotted against $(\rho g)^{3/4}(EH)^{1/4}\xi_0^{3/2}$ (eq 9a) for seven different elastic films (1.8 $\mu$m to 14.4 $\mu$m) (D) $F/L$ is plotted against $(\rho g)^{2/3}(EH)^{1/3}\xi_0^{4/3}$ (eq 9b) for the same films as above.

On the other hand, a plot of $F/L$ against $(\rho g)^{2/3}(EH)^{1/3}\xi_0^{4/3}$ for the same films yields a slope of $1.03\pm0.01$, which, in fact is in much better agreement with the theoretically predicted value of 1.1 (equation 9b) than that of the previous plot. The excellent collapse of the load-displacement data, nevertheless, suggests that all the elastic films have very similar Young's modulus.

**2.5. Attraction of Cylinders on Thin Elastic Films.** We studied the attraction of two identical cylinders on three PDMS films of thickness ranging from 5.7 $\mu$m to 12.6 $\mu$m while they rolled and made final contact in perfect alignment on the film (Figures 5 a-b ). The force of attraction between the cylinders increases as their separation distance decreases with concomittant descent of the cylinders in the liquid while still floating atop the film. The results obtained from three different sets of experiments were compared by plotting the normalized descent of the cylinder $\Delta h/h_\infty$ against the normalized separation distance $\alpha \ell$. The $\alpha^{-1}$ values (4.3 mm, 5.5 mm and 6 mm respectively) used for these analyses were obtained from the experiments described in section 2.3 C. Although an excellent collapse of the data was obtained for three different films with each experiment repeated 5 times, the magnitude of the attraction energy is lower than that predicted by equation 7. We discuss the possible origin of this discrepancy in the following section.

At this juncture, we point out that we estimated the gravitational potential energy of the cylinders with the depth $h$ estimated from its base from the undeformed surface of the elastic film (figure 5a), whereas $\xi_0$ is the distance of the contact line where the film meets the cylinder from the undeformed surface. The separation distance between the contact lines of the two cylinders is only slightly greater than that of the contact edges. Fortunately, the errors associated with these approximations are rather small and thus shifting (Figure 5 c) all the minimum values of $\alpha \ell$ to zero introduces negligible error in the estimation of the net attraction energy.

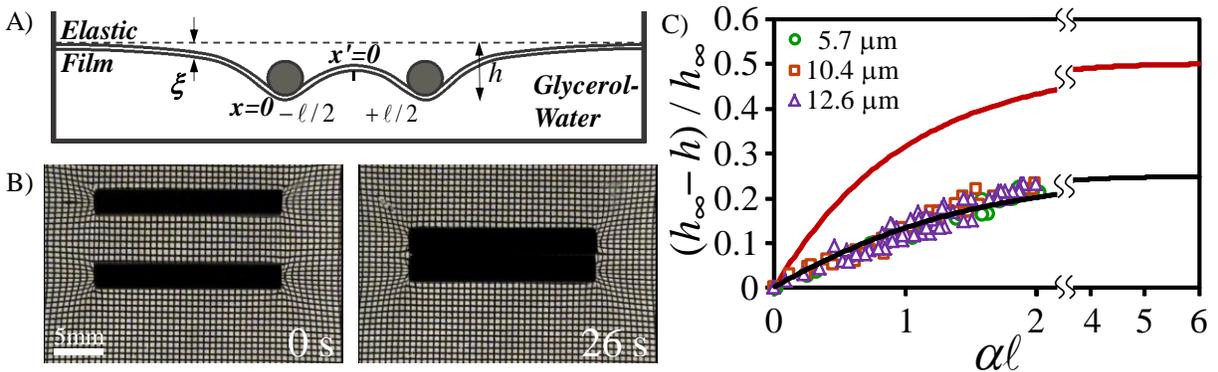

**Figure 5.** (A) Schematic of the side view for the attraction of two PDMS coated steel cylinders on the surface of an elastic film (PDMS 1:1 Sylgard 184 and 186) supported on glycerol-water solution in a polystyrene petri dish. (B) Plan view of the attraction of two steel cylinders from an intermediate separation distance ( 5 mm) to final contact preserving parallel alignment. The wire

mesh lined with the base of the petri dish shows the deformation field of the film. (C) This graph summarizes the non-dimensional descents of the cylinders $\Delta h/h_\infty$ as a function of the non-dimensional distance of separation $\alpha\ell$. The red curve shows the theoretically predicted (Equation 7) energy of attraction in an ideal situation without adhesion hysteresis. The experimental data for the attraction of two cylinders on three different films (5.7, 10.4 and 12.6 µm) show a good collapse but being much lower in magnitude than that predicted from theory. The theoretical black curve accounts for the role of adhesion hysteresis (Equation 12).

## 2.6. Role of Adhesion Hysteresis

Equation 7 is applicable when the cylinders roll freely from a pre-determined distance on the elastic film till the final contact is established. However, rolling hardly occurs freely on any surface. Rolling of a cylinder on a surface can be viewed as the propagation of two cracks, one in the front and the other at its rear edge[18, 19]. The energy to break contact is usually somewhat higher than the energy gained in making contact. Thus, not all of the available energy of attraction is converted to the gravitational potential energy of the cylinders -- some of the energy is stored in stretching the film behind the cylinders. In the absence of a detailed model, we assume that this additional energy is proportional to the square of the descents of the cylinders from the starting position (third term on the RHS in equation 10):

$$U^* = -2m^*g\xi_0 + \frac{\rho g \xi_0^2 L}{\alpha}\left(1 + \frac{\sinh(\alpha\ell)}{2\cosh^2(\alpha\ell/2)}\right) + C(\xi_0 - \xi_{01})^2 \qquad (10)$$

Using the stability condition $\partial U^*/\partial \xi_0 = 0$, we obtain an expression for $\xi_o$ as a function of $\ell$:

$$\xi_0 = \frac{m^*g + C\xi_{01}}{\frac{\rho g L}{\alpha}\left(1 + \frac{\sinh(\alpha\ell)}{2\cosh^2(\alpha\ell/2)}\right) + C} \qquad (11)$$

At this point, we again replace $\xi_o$ with $h(\ell)$ and $\xi_{o1}$ with $h_0$ as the depth of submersion ($h$) of the cylinders into the film for the reasons already discussed above. Evaluation of $C$ from equation 11 yields a modified form of equation 7 that takes into account the role of adhesion hysteresis, where all the parameters can be estimated experimentally as shown below:

$$\frac{h_\infty - h(\ell)}{h_\infty} = \frac{\sinh(\alpha\ell)}{\left\{\dfrac{m^*\alpha}{\rho L(h_\infty - h_0)}\right\}\cosh^2(\alpha\ell/2) + \sinh(\alpha\ell)} \tag{12}$$

### 2.7. Analysis of the Energetics of Attraction

It is evident in figure 5 c that equation 7 over-predicts the descents of the cylinders as a function of the separation distance than what is observed experimentally. When the denominator of equation 12 is calculated with the corresponding value of $m^*\alpha/[\rho L(h_\infty - h_0)]$, the corrected values of the descents of the cylinders are in excellent agreement with those observed experimentally.

Based on the above discrepancy between the experimental results and theoretical predictions of the descends of the cylinders, it is possible to make an approximate estimate of the magnitude of adhesion hysteresis as follows. The main difference comes from the theoretical value of $(h_\infty - h_0)/h_\infty = 0.5$ without adhesion hysteresis and that $[(h_\infty' - h_0)/h_\infty' = 0.25]$ with adhesion hysteresis. The adhesion hysteresis $\Delta W$ (energy/area) is the difference in the energies of adhesion in opening a crack at the trailing edge and closing a crack at the advancing edge of the rolling cylinder. For a cylinder of length $L$ (19 mm) rolling over a distance $\ell$ (8 mm), the energy due to adhesion hysteresis can be represented as: $(\Delta W)\ell L = m^* g(h_\infty - h_\infty')$. From the experimentally estimated value of $(h_\infty - h_\infty')$, the adhesion hysteresis $\Delta W$ is estimated to be about 135 mJ/m$^2$, which we now compare with that obtained from the forced rolling of a cylinder on the surface of a PDMS film.

### 2.8. Adhesion Hysteresis From Forced Rolling

We confirmed the presence of adhesion hysteresis at the interface of the steel cylinder and a PDMS film as follows. After placing a PDMS coated cylinder on the elastic film it was pushed by one end of a tungsten wire spring. As the stage containing the cylinder translated quasi-statically, the cylinder deflected the spring as it rolled on the PDMS film (Figure 6). Three different elastic films (7, 11, 16 μm) were used to perform these experiments. By knowing the spring constant of the wire, the adhesion hysteresis ($\Delta W = F/L$) was estimated[18,19] from the deflection of the spring at the onset of rolling that yielded a value close to 100 mJ/m$^2$ for all the films, which agrees well with the value (135 mJ/m$^2$) reported in section 2.7 above.

**2.9 Attraction of Cylinders on a Hydrogel Coated Elastic Film.** In view of the previous observations, it is transparent that rolling friction impedes the attraction of cylinders beyond only about 3-4 times its diameter. We hypothesized that the cylinders could attract from a much larger separation distance taking lesser amount of time than what is observed with a bare

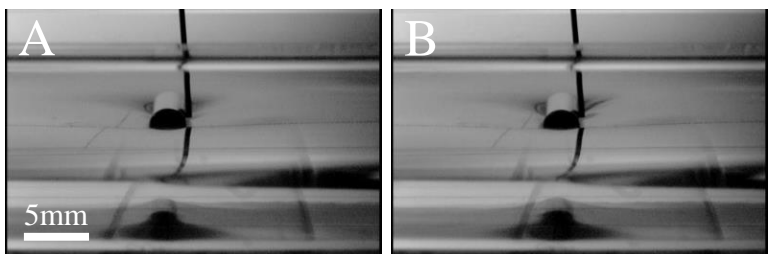

**Figure 6.** Displacement of the end of a tungsten spring wire caused by the rolling of a cylinder on the surface of a 11.2 μm elastic film supported on glycerol water solution. Image (A) shows when the wire just touches the cylinder and image (B) shows the maximum deflection the wire spring as the cylinder continues to roll on the surface.

PDMS film if the impeding friction is reduced or eliminated. In order to test this hypothesis, we designed an experiment in which a thin uniform layer of a low modulus (~ 10 Pa) hydrogel (1.5mm thick) was cross-linked above the thin PDMS film so that the deformation of the PDMS/Gel composite layer provides the energy of attraction, whereas the low friction hydrogel affords this attraction to commence from a large distance (Figure 7 a). What we observe on the surfaces of these supported gel films is the steel cylinders (as well as steel spheres) attract from an initial separation distance of about 6 times the diameter of the particles (Figure 7 b-c), that is almost more than double of what is observed with the cylinders on the bare elastic films. Because of the low friction of the hydrogel film, the cylinders occasionally attain intermediate non-parallel configurations that is similar to that observed with the cylinders on the surface of a liquid[2]. When several particles are released on such a surface, they formed clusters (figure 8).

In order to prove that the long range attraction observed with the hydrogel-PDMS film composite is indeed due to their complementary properties, we performed a control experiment in which the cylinders were placed in close proximity on the surface of a thin hydrogel film cured against a rigid substrate (glass plate or flat base petri dish). Here, either no visible interaction, or sometimes a very weak attraction, was observed between the cylinders or particles, thus suggesting that the long range attraction observed on the composite of the hydrogel and PDMS film is unique. A more detailed study of the interaction of particles on thin hydrogel film on a rigid substrate is reserved for a future in-depth study as the shear deformation in such thin films could give rise to new length scales of attraction and repulsion of particles on its surface.

**2.10. Role of Instability on the Hydrogel Coated Elastic Film.** Soon after the hydrogel films are deposited on the PDMS film, it remained smooth. However, with time, morphological instabilities develop on the surface of the hydrogel (Figure 9 a) that leads to a rough energy terrain with intermittent barriers. When particles are deposited on such a surface they do not necessarily move in straight path; instead they follow corrugated (minimum energy) paths guided by the folds on the surface. Occasionally, the particles get pinned on such a surface at local energy minima, which prevents them from attracting from even a very small separation distance. When many particles are released on such a surface, they formed clusters (Figure 9b) around the edges of the folded structures, which are reminiscent (in a microscopic sense) of the assembly of particles along the defects of liquid crystals.[15]

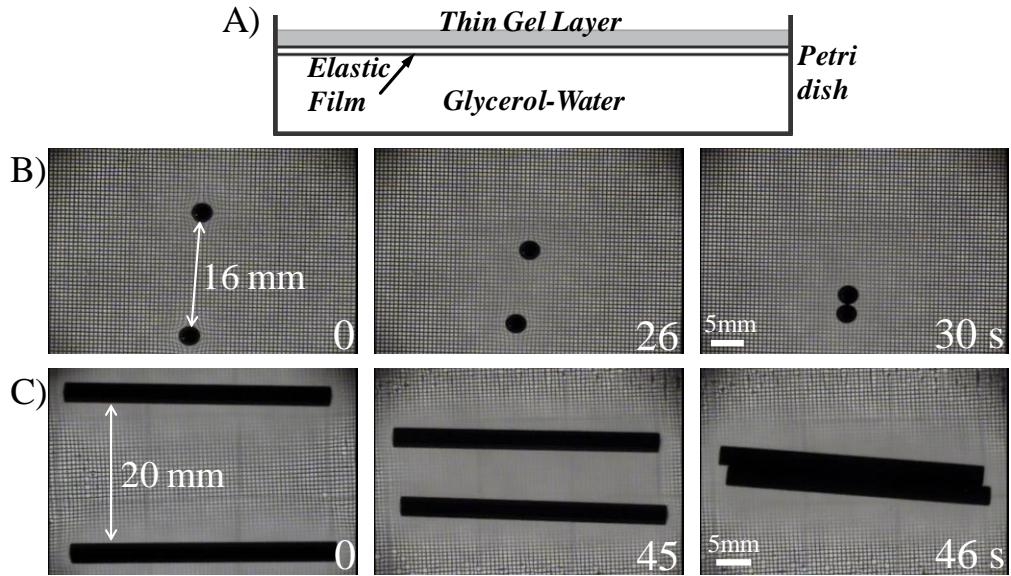

**Figure 7.** (A) Schematic of the thin gel layer (1.5 mm thick, 10 Pa shear modulus) supported on the elastic film (14.4 µm thick) on the pool of liquid. A super-wetting silicone surfactant was added to the gel to promote its spreading on the PDMS film (B) Long range attraction of two steel spheres (diameter 3mm) making final contact on the surface of the gel layer. (C) Long range attraction of two cylinders (length 1.5") approaching each other in a parallel fashion on the similar gel supported on 19.7 µm thick elastic film.

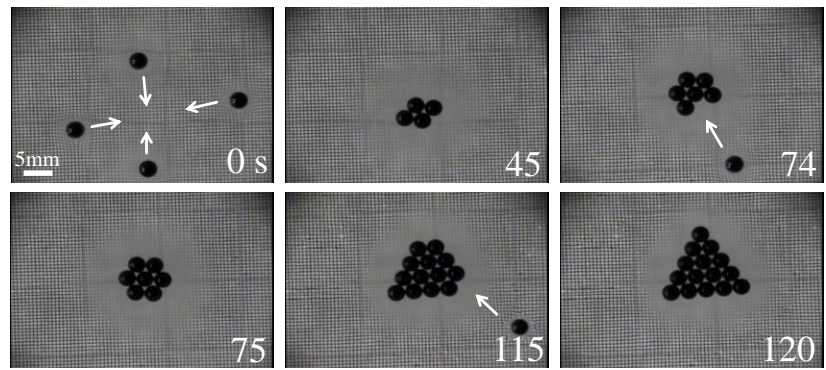

**Figure 8.** Assembly of particles via long range attraction on the surface of a thin hydrogel layer supported on a 19.7 µm elastic film. The steel spheres (diameter 3mm) seek the minimum energy state and move towards the gaps crated by the neighboring spheres. The white arrows indicate the direction of the movement of the spheres.

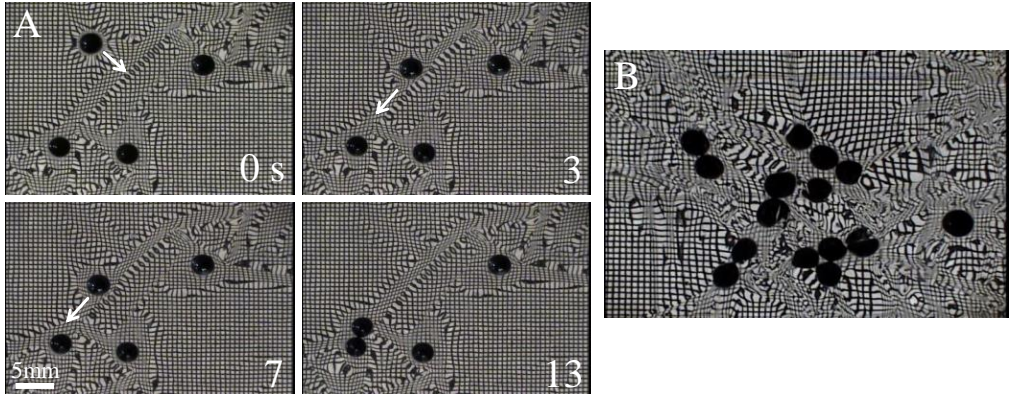

**Figure 9.** (A) Surface instabilities on the hydrogel layer (1.5mm) supported on a 12.6 μm elastic film guide the corrugated paths of particles (steel sphere, diameter 3mm) as indicated by the white arrows, following a minimum energy path. The wire mesh lined with the base of the petri dish helps in the visualization of the instabilities. (B) Several metastable states as created on a hydrogel layer (1mm) supported on the surface of a 7.4 μm elastic film prevent a global clustering of the steel spheres even when the separation distance is small, even though local aggregation of particles are evident near the defects.

## 3. Summarizing The Main Points

As far as we know, this is the first systematic study of the attraction of two cylinders on a surface that preserves its parallel configuration from a modest separation distance all the way to contact. This contrasts with what is observed[2] with the attraction of cylinders on liquid surface, where they attain several intermediate unstable configurations before coming into final contact without necessarily preserving a parallel configuration. The difference in these two types of behaviors lies in the sliding friction between the cylinder and the substrate, a finite value of which stabilizes the parallel configuration on the surface of a thin elastomeric film. The results presented here demonstrate that a thin PDMS film prepared from the mixtures of Sylgard 184 and 186 supported on a pool of glycerol-water solution is a viable way to study interaction of particles mediated by elastic tension and gravity. The energy analysis demonstrates that the energetics of attraction can be adequately quantified by the change in the gravitational potential energy resulting from the descents of the cylinders into the liquid while still floating on the surface. Examination of the profile of the deformed surface in the vicinity of the mid-section of the cylinder provides a slightly more reliable estimate of the characteristic material length scale $\alpha^{-1} = \sqrt{T/\rho g}$ than that estimated from the side view due to the obscurities resulting from folding and wrinkling instabilities of the film. However, the value of the solid surface tension of the film ($56 \pm 27$ mN/m) obtained from the extrapolation of $\alpha^{-1}$ to zero film thickness is unreliable due to large standard deviation. This result, however, contrasts a recent report[25] where significantly higher values (100 – 200 mN/m) of solid surface tension for PDMS was obtained from the deformations of thin films caused by liquid drops. The observations of the exponential profile of the thin film PDMS when deformed by a cylinder as well the results of the indentation experiments are all consistent with a linear elasticity model. Hence, no attempt was made to invoke a non-linear elasticity model in all the subsequent analysis of the data presented in this study.

Although the change of the gravitational potential energy of the cylinder provides an easy option to estimate the energy of attraction of the cylinders, it is found that not all of the available energy is converted to the gravitational potential energy. This discrepancy can be ascribed reasonably to the hysteresis of adhesion due to rolling of the cylinders on the elastic film, which in turn converts some of the energy to stretching the film behind the cylinders as they approach each other. This conjecture has been supported by both an energy analysis accounting for hysteresis and an experiment that provided direct evidence of its presence at the cylinder-film interface. While performing the experiments involving the attraction of the cylinders on the elastic film, we frequently noticed that when the initial separation distance is much larger than the decay length $\alpha^{-1}$, they approach towards each other initially by a slight amount, but then no perceptible movement is observed. This is an evidence of the adhesion hysteresis providing a threshold force that prevents the rolling of the cylinders on a thin elastic film. Then again, we made some infrequent observations in which the cylinders approach towards each other very slowly over a long time (> 30 min) and come into contact starting from a rather large separation distance (~ 1.5 cm) , thus suggesting that hysteresis may relax with time. However, a further in-depth study is needed to characterize the origin of the hysteresis, as it is not entirely clear, at present, if this hysteresis only provides a threshold force or it relaxes with time. In particular, the roles of the weak bonds between the cylinder[19] and the elastic film or formation of the wetting ridges[26] at the trailing edge of the cylinders rolling on the film need to be investigated.

The role of rolling resistance due to adhesion hysteresis, motivated us to carry out a new experiment with which to observe what happens when such a frictional resistance is eliminated. Indeed, with the deposition of a low friction hydrogel layer atop the thin elastomeric film led to a much longer range attraction of the cylinders (as well as solid spheres) than what was observed with the PDMS film alone. We reserve further analysis of the distance dependent attraction energy in such a system for future. Such a composite film, however, became the ground for more fascinating experiments involving the assembly of multiple particles on its surface. A potentially important observation is that morphological instabilities on the hydrogel layer may induce creation of metastable states leading to a mechanically tunable rough energy terrain. We may expect that more interesting studies could be performed with such a surface exhibiting rugged energy landscapes with which to perform mechanical computation of the paths that the particles would follow to reach a global energy minimum state.

The model system in this study and a 2d analysis used to describe the phenomena reported in this work are deceptively simple. The real situation is somewhat more complex. For example, surrounding the out-of-plane stretched region brought about by the weight of the cylinder lie a region where compressive stress develops in the film. In this compressed region, bending of the film also plays a role, which induces wrinkling and the folding instabilities that are evident near the two poles of the cylinders (Figure 1). Even beyond the out-of-plane stretched region that we subscribe as the "active zone" lie a region extending all the way to the wall of the container, where the film undergoes an in-plane stretching that accommodates the liquid displaced from the out-of plane stretched zone. While the resulting gravitational head of the liquid near the walls may be ignored while accounting for the profile in the out-of-plane stretched region, it should play a role in defining the state of the film in the in-plane stressed region. Full understanding of the problem, therefore, would require a rigorous analysis of the mechanics of thin film (e.g. extending the type of analysis reported in reference 27) while sacrificing the simplicity used here. Nevertheless, the success of the approximate analysis to account for the main observations reported here may be motivational in terms of the reasonableness of the approximations that need to be made in developing a more rigorous 3d analysis of the problem.

## 4. Conclusion

Our system comprising of a thin elastic film supported on a pool of liquid allows the study of interaction of particles on its surface, in which a two dimensional energy minimization captures the main underlying physics of the problem. An advantage of studying interactions on thin elastic film is that both the strength and the range of interactions can be easily controlled by the thickness and the elasticity of the film. The work points out the eminent role of adhesion hysteresis between the particle and the film that impedes long range attraction to an appreciable degree. This problem can, however, be alleviated by supporting a thin layer hydrogel on the elastic film, where the low friction of the hydrogel and the deformability of the membrane provide complementary properties with which a long range attraction can be studied. Modification of the property of the hydrogel (including its morphological instability) and that of the elastic membrane could vastly extend the range of studies involving elasto-capillarity thus enriching the scope of new physics to be discovered.

## 5. EXPERIMENTAL DETAILS

**5.1. Materials.** Stainless steel cylinders (Length ¾", Diameter 1/8", Density 7.8g/cm$^3$) were used for all the experiments in this study that were treated with trimethylsiloxy-terminated polydimethylsiloxane (DMS T-22, M.W. 9430; Gelest Inc.) for 24 h in the oven followed by an oxygen plasma cleaning for an hour. This is a modified method of the treatment explained by Krumpfer et al.[28] The contact angle of water on such a treated cylinder was found to be around 90°. Steel balls (E52100 alloy steel or SS316, density 7.8g/cm$^3$, McMaster Carr) were used as is. A stainless steel (SS316) wire cloth (opening size 0.015 in., wire diameter 0.010 in., McMaster Carr), was lined with the base of the glass cell to observe the deformations in the gel as the particles interacted for the plan view images of the experiments.

**5.2. Preparation of the Elastic Film on a Pool of Liquid.** A 1:1 solution (density $\rho$ 1.13 g/cc, 20°C) of glycerol (Fisher Chemical) and deionized (DI) water was degassed for 30 min using a vacuum pump (Welch Duo-Seal, Model no. 1402). Sylgard 184 and Sylgard 186 (Dow Corning®) were mixed in 1:1 ratio (the amount of crosslinker added to the mixture was 10% of the weight of the base polymer) that was degassed for 10 min under high vacuum. Such a combination of the polymers was chosen to ensure that the film has a high tear strength as well as a moderately high elastic modulus[24]. All the samples were prepared in square polystyrene petri dishes (VWR, 100mm ×100mm ×10mm). The petri dishes were filled with the glycerol-water solution to about half of their height. Different quantities of the PDMS mixture were gently released on the surface of the liquid with the help of micropipettes and the added amount was weighted using a sensitive balance. The samples were allowed to stand for 15 min that allowed the spreading of the PDMS mixture on the liquid surface following which they were carefully placed inside an oven and cured for 90 min at 75° C. Using the known cross sectional area of the dish and the weight of the polymer added, the thicknesses of the elastic films (2 - 17 $\mu$m) were estimated. The cured samples were cooled before using them for further experiments. The time duration from the preparation of the samples to further experiments was >2 hrs. For the purpose of plan viewing of the samples, some round petri dishes (VWR, 100mm diameter, 10 mm high) containing the samples were used occasionally.

**5.3. Estimation of the Decay Length ($\alpha^{-1}$).** The decay lengths of the different thickness elastic films were estimated using two different techniques. In the first technique (discussed in Section 2.3 A), a cylinder was placed on the surface of the film and the deformed profile was imaged using a microscope (Infinity) equipped with a Charge Coupled Device camera (MTI, CCD-72) from the side through the transparent wall of the polystyrene petri dish. The calibration factor was obtained from the image itself from the known diameter of the cylinder.

In the second technique (discussed in Section 2.3 B), a linear trace of ink was made by a glass slide whose edge was inked with a black water-based marker (Crayola). A cylinder was placed at the centre of the petri dish such that the ink trace was at its mid section. The deformed profile of the ink was then imaged using a regular camera (Samsung, angle of lens at 75 $^0$ from the horizontal surface of the film) (figure 2 a). Prior to performing the final experiments with the elastic films, we verified the appropriateness of this method and estimating the appropriate calibration factors by taking an image of a graph having an pre-designed exponential profile. On each image, a horizontal line was drawn parallel to the undeformed portions of the ink line at the two sides of the cylinder. The composite image with the line was then analysed using ImageJ where the vertical distance of the deformed ink trace was calculated from this horizontal line as a reference. In all experiments, it was ensured that vibration has no significant effects in the measurements.

**5.4. Attraction of Cylinders on the Elastic Film.** Experiments involving the attraction of cylinders were carried out on three different PDMS films of thicknesses 5.7 $\mu$m, 10.4 $\mu$m and 12.6 $\mu$m. Two identical PDMS coated stainless steel cylinders (Length ¾", Diameter 1/8") were placed on the surface of the film such that they were in perfect alignment. With the help of two pairs of tweezers, they were then separated to a considerable distance (several times the diameter of the cylinders) and then brought back slowly to about 1 cm separation till they started to attract each other. We recorded the attraction as the cylinders rolled towards each other and descended in the liquid with a CCD camera, MTI CCD-72. The separation distance between the cylinders and their descents ($h$, the vertical distance of the base of the cylinder from the undeformed surface of the gel) were noted using the ImageJ software.

**5.5. Adhesion Hysteresis: Rolling of Cylinder on Elastic Film.** A cylinder was placed on the sample containing the elastic film that was placed on the X-Y manipulator stage. A tungsten wire (diameter 0.02", SPI) of a known spring constant was fixed above the cylinder so that it just touched one edge of the cylinder (figure 6 a). The stage was moved quasistatically so that the cylinder rolled on the elastic film while displacing the spring wire laterally. From the maximum deflection of the wire, the adhesion hysteresis was estimated using the equation[19] $\Delta W = F/L$ as shown in section 2.8. The spring constant of the wire was determined from the deflection of its one end by hanging a known weight from it. The image analysis was performed in ImageJ.

**5.6. Hydrogel Coated Elastic Film.** In a clean glass jar, N-(hydroxymethyl)-acrylamide (48% solution in water, Sigma Aldrich) and Deionized water were mixed to prepare a 3.2% (w/w) of the monomer in the solution that was followed by purging it with ultrapure nitrogen gas for 30 minutes while stirring it constantly. This was followed by stirring in 0.25 wt% Potassium Persulphate (99.99% trace metals basis, Sigma Aldrich) for 10 min. Few drops of a surfactant (Q2-5211 Superwetting agent, Dow Corning ®) was stirred in following the addition of 0.3 wt% N,N,N′,N′- tetramethylethylenediamine (TEMED, ≥99.5%, purified by redistillation, Sigma Aldrich). The surfactant ensured that the gel solution spreaded completely on the surface of the PDMS thin film, which afforded preparation of a thin gel layer. The shear modulus of such a gel is about 10 Pa[16]. The experiments involving the long-range attraction of particles on the Gel/PDMS composite were performed after 1 h of curing the gel. Morphological instabilities developed on some of the gels that were allowed to stay in the ambient condition for a much longer time (>2 h).

**Acknowledgement** An exchange with L. Mahadevan during the early phase of this study was motivational. We greatly benefited from the pedagogical exchanges with Professors A. Jagota, Y. Pomeau and K-T. Wan. While we are grateful to the above colleagues for their insightful comments, we do not imply that they are in total agreement with our analysis. We take sole responsibility for any mistakes incurred with the current study. We also thank Charles Extrand for some valuable comments.

**Appendix**

**Justification of Equation 1**

The constancy of the tension of the film is related to the fact that there is no interfacial shear stress to balance the gradient of stress along the deformed arc length of the film. If a rectangular strip anchored on both ends is depressed in its middle by the weight of a cylinder, then the entire film will have uniform tension of magnitude, $EHu_o/L$, where $u_o$ is the total deformation and $L$ is the total length of the film. In our case, however, because of lateral constraints, the film cannot deform all the way to the length scale of the container. There will be a cut-off length ($L^*$), up to which the tension will be more or less uniform and beyond which the tension should be vanishingly small. Since, $u_o \sim (1/2)\int_0^\infty \xi_x^2 dx$; the elastic tension is: $T_E \sim (EH/2L^*)\int_0^\infty \xi_x^2 dx \sim EH\xi_o^2 \alpha/4L^*$.

The total energy functional comprising of the gravitational potential energy of the liquid and the stretching and the surface energies of the film on both sides of a cylinder resting on its surface is thus given as follows:

$$U_f = L\int \rho g \xi^2 dx + T_E L\int \xi_x^2 dx + \gamma L \int \xi_x^2 dx \tag{A1}$$

A functional derivative of $U_f$ yields:

$$\frac{\delta U_f}{\delta \xi} = 2L\int \left(\rho g \xi - (\gamma + T_E)\xi_{xx}\right) dx \tag{A2}$$

Setting $\delta U_f / \delta \xi = 0$ and denoting the $\gamma + T_E$ as $T$, we obtain the desired differential equation the solution of which yields the exponential profile of the film.

**Discussion of Equations 9:**

Substituting $\xi = \xi_0 e^{-\alpha x}$ in equation A4 (ignoring the role of surface tension $\gamma$) yields:

$$U_f / L = \rho g \xi_o^2 / 2\alpha + 3\mu H \xi_o^4 \alpha^2 / (16 L^*) \tag{A3}$$

At this point, we make two bold (ad hoc) assumptions and test their validities experimentally. The first one is to assume that $L^*$ is on the order of $1/\alpha$, but somewhat larger than the latter. With this assumption, the minimization of $U_f/L$ with respect to $\alpha$ yields $\alpha \sim \left(8\rho g / 3EH\xi_o^2\right)^{1/4}$.

Eliminating $\alpha$ in equation A4 and taking the derivative of the energy with respect to $\xi_0$, we obtain equation (9A) in the text. The experimental data are, however, more consistent with the strain scaling as $\xi_0 \alpha$, which yields equation (9B) of the text. The change of the power index could be the result of the finite contact area between the cylinder and the film due to adhesion that we have neglected so far.